\documentclass[10pt]{article}
\usepackage{graphicx}
\usepackage{amssymb}
\usepackage{amsmath}
\usepackage{color}
\numberwithin{equation}{section}
\numberwithin{figure}{section}
\numberwithin{theorem}{section}
\numberwithin{teorema}{section}

\def\be{\begin{equation}}
\def\ee{\end{equation}}
\def\bc{\begin{center}}
\def\ec{\end{center}}

\title{\bf Equilibrium statistical mechanics on correlated random graphs}
\author{Adriano Barra\ $^1$, Elena Agliari\ $^2$}
\begin{document}
\date{}
\maketitle

\begin{center}
{\small
\vskip-0.5cm


\footnote{e-mail:{\tt  adriano.barra@roma1.infn.it}} Dipartimento
di Fisica, Sapienza Universit\`a di Roma (Italy) \vskip-0.5cm GNFM
 Gruppo Nazionale per la Fisica Matematica \vskip-0.5cm

\footnote{e-mail:{\tt  elena.agliari@fis.unipr.it}} Dipartimento di
Fisica, Universit\`a di Parma (Italy) \vskip-0.5cm INFN, Gruppo Collegato di Parma (Italy) \vskip-0.5cm Theoretische Polymerphysik, Albert-Ludwigs-Universit\"{a}t, Freiburg (Germany)}
\end{center}


{\small \bf------------------------------------------------------------------------------------------------------------}

\vskip-1cm
{\small

{\bf Abstract.}

Biological and social networks have recently attracted enormous
attention between physicists. Among several, two main aspects may
be stressed: A non trivial topology of the graph describing the
mutual interactions between agents exists and/or, typically, such
interactions are essentially (weighted) imitative. Despite such
aspects are widely accepted and empirically confirmed, the schemes
currently exploited in order to generate the expected topology are
based on a-priori assumptions and in most cases still implement
constant intensities for links.
\newline
Here we propose a simple shift $[-1, +1]\to[0, +1]$ in the
definition of patterns in an Hopfield model to convert frustration
into dilution: By varying the bias of the pattern distribution,
the network topology -which is generated by the reciprocal
affinities among agents (the Hebbian kernel)- crosses various well
known regimes (fully connected, linearly diverging connectivity,
extreme dilution scenario, no network), coupled with small world
properties, which, in this context, are emergent and no longer
imposed a-priori.
\newline
The model is investigated at first focusing on these topological
properties of the emergent network, then its thermodynamics is
analytically solved (at a replica symmetric level) by extending
the double stochastic stability technique, and presented together
with its fluctuation theory for a picture of criticality: both a
statistical mechanics and a topological phase diagrams are
obtained.
\newline
Overall the picture depicted from statistical mechanics is quite
intuitive: at least at equilibrium, dilution (of whatever kind)
simply decreases the strength of the coupling felt by the spins,
but leaves the paramagnetic/ferromagnetic flavors unchanged.
\newline
The main difference with respect to previous investigations and a
naive picture is that within our approach replicas do not appear:
instead of (multi)-overlaps as order parameters, we introduce a
class of magnetizations on all the possible sub-graphs belonging
to the main one investigated: As a consequence, for these objects
a closure for a self-consistent relation is achieved.
%
%
}


\cleardoublepage


\section{Introduction to social and biological networks}

The paper is organized as follows:

In this section we briefly introduce the reader to the state of
the art in the applications of this model to investigation of
collective effects in social and biological networks, then, in
section $2$, we present the model itself with all the related
definitions. Section $3$ deals with the topological analysis:
Techniques from graph theory are the tools. Section $4$ deals with
the thermodynamical analysis: techniques from statistical
mechanics are the tools. In section $5$ we present our discussion
and outlooks.

Starting with a digression on social sciences, since the early
investigations by Milgram \cite{milgram}, several efforts have
been made to understand the structure of interactions occurring
within a social system. Granovetter defined this field of science
as "a tool for linking micro and macro levels of sociological
theories" \cite{grano1} and gave fundamental prescriptions; in
particular, he noticed that the stronger the link between two
agents and the larger (on average) the overlap among the number of
common nearest neighbors, i.e. high degree of cliqueness.
Furthermore he noticed that weak ties play a fundamental role
acting as bridges among sub-clusters of highly connected
interacting agents \cite{grano1,grano2,weakeco}. As properly
pointed out by Watts and Strogatz \cite{watts}, from a topological
viewpoint, the simplest Erdos-Renyi graphs \cite{bollo} is unable
to describe social systems, due to the uncorrelatedness among its
links, which constraints the resulting degree of cliqueness to be
relatively small \cite{barabasi}. Through a mathematical technique
(rewiring), they obtained a first attempt in defining the so
called "small world" graph \cite{watts2}: when trying to implement
statistical mechanics on such a topology their network has been
essentially seen as a chain of nearest neighbors overlapped on a
sparse Erdos-Renyi graph \cite{ton2,marti}. As the former can be
solved via the transfer matrix, the latter via e.g. the replica
trick, the model was already understood even from a statistical
mechanics perspective (without introducing here a discussion on
possible
 replica symmetry breaking  in complex diluted systems
\cite{monta,ioepl}).
\newline
Coupled to topological investigations, even the analysis of the
kind of interactions (still within a "statistical mechanics
flavor") started in the past decades in econometrics and, after
McFadden described the discrete choice as a one-body theory with
external fields \cite{fadden}, Brock and Durlauf went over and
gave a clear positive interaction strength to social ties
\cite{durlauf1,durlauf2}.
\newline
Even thought clearly, as discussed for instance in \cite{epl}, the
role of anti-imitative actions is fundamental for collective
decision capabilities, the largest part of interactions is
imitative and this prescription will be followed trough the paper.

Somewhat close to social breakthrough, after the revolution of
Watson and Crick, biological studies in the past fifty years gave
raise to completely new field of science as genomics
\cite{franco}, proteinomics \cite{gavin} and metabolic network
investigations \cite{enzo}  which ultimately  are  strongly based
on graph theories\footnote{It is in fact well established that
complex organisms share roughly the same amount of genes with
simpler ones. As a result the failure of a purely reductionism
approach (more genes $\to$ more complexity) seems raising and
interest in their connections, their network of exchanges, is
enormously increasing.} \cite{barabasi2}. Furthermore graph
structure appears at various levels, i.e. in matching epitopal
complementary among antibodies giving raise to the so called
"Jerne network" \cite{jerne}\cite{parisi} for the immune system
\cite{baglia,agliba}, or at even larger scales of the biological
world: from the so far exploited micro and meso, to such a macro
as virus spreading worldwide \cite{vespignani}, food web
\cite{sole}, and much more \cite{guido}.
\newline
In these contexts, surely there is a disordered underlying
structure, but thinking at it as "completely random" is probably a
too strong  simplifying assumption. One of the strongest starting
point when dealing with random coupling is their independence: for
example Blake pointed out \cite{blake} that exons in haemoglobin
correspond both to structural and functional units of protein,
implicitly suggesting a not null level of correlation among the
"randomness" we have to deal with when trying a statistical
mechanics approach. Not too different is the viewpoint of Coolen
and coworkers \cite{tonBio1}\cite{tonBio2}.

\bigskip

From a completely different background, last step in this
introduction is presenting the Hopfield model \cite{hopfield},
which, instead, is the paradigmatic model for  neural networks.
Even though apparently far from topology investigations, in the
Hopfield model there is a scalar product among the bit strings
(the Hebbian kernel \cite{hebb}): despite fully connected, the
latter can be seen as a measure of the strength of the ties (which
in that context must be both positive and negative as, in order to
share  statically memories over all neurons \cite{amit}, it must
use properties of spin glasses
\cite{barra1}\cite{barraguerra}\cite{MPV} as the key for having
several minima in the fitness landscape). By varying tunable
parameters (level of noise and amount of storage memories) the
Hopfield model displays a region where is paramagnetic, a region
where is a spin glass and a region where is a "working memory"
\cite{AGS1}\cite{AGS2}.
\newline
We are ready to introduce our starting idea: what happens if
instead of using positive and negative values for the coupling in
the Hebbian kernel of the Hopfield model, we use positive and null
values?
\newline
We want to show that, even in this context, by varying the tunable
parameters, we recover several topologies (on which ferromagnetic
or paramagnetic behaviors may arise): fully connected scenario
weighted and un-weighted, Erd\"{o}s-R\'{e}nyi graphs, linearly
diverging connectivity, extreme dilutions, small world features,
fully disconnected (that is no edges at all).
\newline
Despite a rich plethora of phenomena in graph theory is obtained,
from equilibrium statistical mechanics perspective we find that
all these networks behave not drastically differently, relating
strong differences in dynamical features (in agreement with
intuition), on which we plan to investigate soon.


\section{The model: Definitions}\label{sec:model}

Let us consider $V$ agents $ \pm1 \ni \sigma_i, i \in (1,...,V)$.
In social framework (e.g. discrete choice in econometrics) for
example $\sigma_i=+1$ means that the $i^{th}$ agent agrees a
particular choice (and obviously disagreement in the $-1$ case).
In biological networks, $i$ may label a Kauffman gene (assuming
undirected links) or a Jerne lymphocytes in such a way that
$\sigma_i=+1$ represents expression or firing state respectively, while quiescence is assumed when $\sigma_i = -1$.
\newline
The influence of  external stimuli, representing e.g. medias in
social networks or environmental variations imposing phenotypic
changes via gene expression in proteinomics or viruses in immune
networks, can be encoded by means of a one-body Hamiltonian term
$H= \sum_i^V h_i \sigma_i$, with $h_i$ suitable for the particular
phenomenon (as brilliantly done by McFadden intro the first class
 of problems \cite{fadden}\cite{barPLgallo},
  Eigen in the middle \cite{peliti} and Burnet in the last class \cite{burnet}\cite{bagliadyn}). As for collective influences
among agents modeling is by far harder.

In the model we are going to develop, each agent $i\in(1,...,V)$
is endowed with a set of $L$ characters denoted by a binary string
$\xi_i$ of length $L$. For example, in social context this string
may characterize the agent and each entry may have a social
meaning (i.e. $\xi_i^{\mu=1}$ may take into account an attitude
toward the opposite sex such that if $\xi_i^{\mu=1}=1$, $\sigma_i$
likes the opposite sex, otherwise if $\xi_i^{\mu=1}=0$; in the
same way $\xi_i^{\mu=2}$ may accounts for smoking and so on up to
$L$). In gene networks the overlap among bit strings may offer a
measure of phylogenetic distance while in immunological context
may offer the affinity matrix built up by strings standing for the
antibodies (and anti-antibodies) produced by their corresponding
lymphocytes.
\newline
Now we want to associate a weighted link among two agents by
comparing how many similarities they share (note that $0-0$ does
not contribute in this scheme, but only $1-1$), namely
\begin{equation} \label{eq:J_ij}
J_{ij}= \sum_{\mu=1}^L \xi_i^{\mu}\xi_j^{\mu}.
\end{equation}
This description naturally
leads to the emergence of a hierarchical partition of the whole
population into a series of layers, each layer being characterized
by the sharing of an increasing number of characters.  Of course,
group membership, apart from defining individual identity, is a
primary basis both for social and biological interactions and
therefore acquaintanceship. As a result, the interaction strength
between individual $i$ and $j$ increases with increasing
similarity.

Hence, including both terms (one-body and two-bodies) the model we
are describing reads off as \be\label{modello} H_V(\sigma; \xi) =
\frac1V \sum_{i<j}^V J_{ij}(\xi) \sigma_i \sigma_j + \sum_i^V
h_i\sigma_i, \ee formally identical to the Hopfield model.
\newline
The string characters are randomly distributed according to
\begin{equation} \label{eq:character_distribution}
P(\xi_i^{\mu}=+1)=  \frac{1+a}{2}, \;\; P(\xi_i^{\mu}= 0)=
\frac{1-a}{2},
\end{equation}
in such a way that, by tuning the parameter $a \in [-1,+1]$, the
concentration of non null-entries for the $i$-th string $\rho_i =
\sum_{\mu} \xi_i^{\mu}$ can be varied. When $a \to -1$ there is no
network and we are left with a non interacting spin system, while
when $a \to +1$ we have that $J_{ij}=L$ for any couple and
(renormalization trough $L^{-1}$ apart) we recover the standard
Curie-Weiss model.

Further, when $a \neq 0$ the pattern distribution is biased,
somehow similarly to the correlations investigated by Amit and
coworkers in neural scenarios \cite{AGS3}. Moreover, from
Eq.$(2.3)$ we get  $\langle \xi_i^{\mu}\xi_i^{\nu} \rangle =
((1+a)/2)[\delta_{\mu \nu}+((1+a)/2)(1-\delta_{\mu \nu})]$, -
apart $a=0$ which reduces to completely uncorrelated patterns.

As we will see, small values of $a$ give rise to highly
correlated,  diluted networks, while, as $a$ gets larger the
network gets more and more connected and correlation among links vanishes.

Even though the theory is defined at each finite $V$ and $L$, as
standard in statistical mechanics, we are interested in the large
$V$ behavior (such that, under central limit theorem permissions, deviations from
averaged values become negligible and the theory predictive). To
this task we find meaningful to let even $L$ diverge linearly with
the system size (to bridge conceptually to high storage neural
networks), such that $\lim_{V\to \infty}L/V = \alpha$ defines
$\alpha$ as another control parameter. Finally, since we are
interested in the regime of large $V$ and large $L$ we will often
confuse $V$ with $V-1$ and $L$ with $L-1$.

\section{The emergent network}

The set of strings 
$\{ \xi_i^{\mu} \}_{ i=1,...,V; \mu=1,...,L }$ together with the
rule in eq. (\ref{eq:J_ij}) generates a weighted  graph
$\mathcal{G}(V,L,a)$ describing the mutual interactions among
nodes. The following investigation is just aimed at the study of its topological features, which, as well known, are intimately connected with
the dynamical properties of phenomena occurring on the network itself (e.g. diffusion
\cite{ben,AB,ABCN}, transport \cite{TC,ACV}, critical properties
\cite{BC,ACD}, coherent propagation \cite{ABM}, relaxation
\cite{mircea}, just to cite a few).
We first focus on the topology neglecting the role of weights and
we say that two nodes $i$ and $j$ are connected whenever $J_{ij}$
is strictly positive; disorder on couplings will be addressed in
Sec.~\ref{sec:Couplings}

It is immediate to see that the number $\rho$ of non-null (i.e.
equal to $1$)  entries occurring in a string $\xi$ is
Bernoulli-distributed, namely
\begin{equation} \label{eq:rho}
P_1(\rho;a,L) = \binom{L}{\rho} \left( \frac{1+a}{2} \right)^{\rho}
\left( \frac{1-a}{2} \right)^{L-\rho},
\end{equation}
with average and variance, respectively,
\begin{eqnarray}  \label{eq:rho_av}
\bar{\rho}_{a,L} = \sum_{\rho=0}^L \rho P_1(\rho; a, L) = \left( \frac{1+a}{2} \right)L,\\
 \label{eq:rho_var}
\sigma^2_{a,L} = \overline{\rho^2}_{a,L}  - \bar{\rho}_{a,L}^2 = \left( \frac{1-a^2}{4} \right) L.
\end{eqnarray}
Moreover, the probability that a string is made up of null entries only is $\prod_{\mu=1}^L P(\xi_i^{\mu}=0) = [(1-a)/2]^L$, thus, since we are allowing repetitions among strings, the number of isolated nodes is at least $V[(1-a)/2]^L$.

Let us consider two strings $\xi_i$ and $\xi_j$ of length $L$, with $\rho_i$ and $\rho_j$ non-null entries, respectively. Then, the probability $P_{\mathrm{match}}(k;\rho_i,\rho_j,L)$ that such strings display $k$ matching entries is
\begin{equation} \label{eq:fond}
P_{\mathrm{match}}(k;\rho_i,\rho_j,L) = \frac{ \binom{L}{k}  \binom{L-k}{\rho_i-k} \binom{L-\rho_i}{\rho_j-k} }{ \binom{L}{\rho_i} \binom{L}{\rho_j} },
\end{equation}
which is just the number of arrangements displaying $k$ matchings
over the number of all possible arrangements. As anticipated, for
two agents to be connected it is sufficient that their coupling
(see eq. (\ref{eq:J_ij})) is larger than zero, i.e. that they
share at least one trait. Therefore, we have the following link
probability
\begin{equation} \label{eq:p_link}
P_{\mathrm{link}}(\rho_i,\rho_j, L) = \sum_{k=1}^L P_{\mathrm{match}}(k; \rho_i,\rho_j,L) = 1 - P_{\mathrm{match}}(0; \rho_i,\rho_j, L) = 1 - \frac{(L- \rho_i)! (L-\rho_j)!}{L! (L - \rho_i - \rho_j)!}.
\end{equation}
The previous expression shows that, in general,  the link
probability between two nodes does depend on the nodes considered
through the related parameters $\rho_i$ and $\rho_j$: When
$\rho_i$ and $\rho_j$ are both large, the nodes are likely to be
connected and vice versa. Another kind of correlation, intrinsic
to the model, emerges due to the fact that, given $\xi_i^{\mu}=1$,
the node $1$ will be connected with all strings with non-null
$\mu$-th entry; this gives rise to a large (local) clustering
coefficient $c_i$ (see section \ref{ssec:SW}). Such a correlation
vanishes when $a$ is sufficiently larger than $-1$, so that any
generic couple has a relative large probability to be connected;
in this case the resulting topology is well approximated by a
highly connected, uncorrelated (Erd\"{o}s-Renyi) random graph.
Moreover, when $a \to +1$ we recover the fully-connected graph.

Finally, it is important to stress that, according to our
assumptions, repetitions among strings are allowed and this,
especially for finite $L$ and $V$, can have dramatic consequences
on the topology of the structure. In fact, the suppression of
repetitions would spread out the distribution $P_1(\rho; a, L)$,
allowing the emergence of strings with a large $\rho$ (with
respect to the expected mean value $L(1+a)/2$); such nodes,
displaying a large number of connections, would work as hubs.
On the other hand, recalling that the  number of couples
displaying perfect overlapping strings is $\sim V^2/2^{L}$, we
have that in the thermodynamic limit and $L$ growing faster than
$\log V$, repetitions among strings have null measure.


\subsection{Degree distribution} \label{ssec:degree}

We focus the attention on an arbitrary string $\xi$ with $\rho$
non-null  entries and we calculate the average probability
$\bar{P}_{\mathrm{link}}(\rho;a) $ that $\xi$ is connected to
another generic string, which reads as
\begin{eqnarray} \label{eq:prob2}
\nonumber
\bar{P}_{\mathrm{link}}(\rho;a) &=& \sum_{\rho_i=0}^L P_1(\rho_i; a, L) P_{\mathrm{link}}(\rho,\rho_i;L)  \\
&=& 1 - \left( \frac{1-a}{2} \right)^L  \left( 1 + \frac{1+a}{1-a} \right)^{L-\rho} =  1 - \left( \frac{1-a}{2} \right)^{\rho}.
\end{eqnarray}
This result is actually rather intuitive as it states that, in
order to be linked to $\xi$, a generic node has to display at
least a non-null entry corresponding to the $\rho$ non-null
entries of $\xi$. Notice that the link probability of
 eq. (\ref{eq:prob2}) corresponds to a mean-field approach where we treat all the remaining nodes  in the average; accordingly, the
degree distribution $P_{\mathrm{degree}}(z;  \rho, a , V)$ for $\xi$
gets
\begin{equation} \label{P_degree}
P_{\mathrm{degree}}(z; \rho, a, V) = \binom{V}{z} \left[ 1 - \left(
\frac{1-a}{2}  \right)^{\rho} \right]^z   \left( \frac{1-a}{2}
\right)^{\rho(V-z)}.
\end{equation}
Therefore, the number of null-entries  controls the
degree-distribution of the pertaining node: A large $\rho$ gives
rise to narrow (i.e. small variance) distributions peaked at large
values of $z$. Notice that $\bar{P}_{\mathrm{link}}(\rho;a)$ and,
accordingly, $P_{\mathrm{degree}}(z;  \rho, a, V)$ are independent
of $L$.

More precisely, from eq.~(\ref{P_degree}), the average degree for a string
displaying $\rho$ non-null entries is
\begin{equation}\label{eq:average_degree_rho}
\bar{z}_{\rho} = V \left[ 1 -  \left( \frac{1-a}{2}  \right)^{\rho} \right],
\end{equation}
while the pertaining variance is
\begin{equation}\label{eq:variance_rho}
\sigma^2_{\rho} = V \left[ 1 -  \left( \frac{1-a}{2}  \right)^{\rho} \right] \left( \frac{1-a}{2}  \right)^{\rho}.
\end{equation}

\begin{figure}[tb]
\includegraphics[height=100mm]{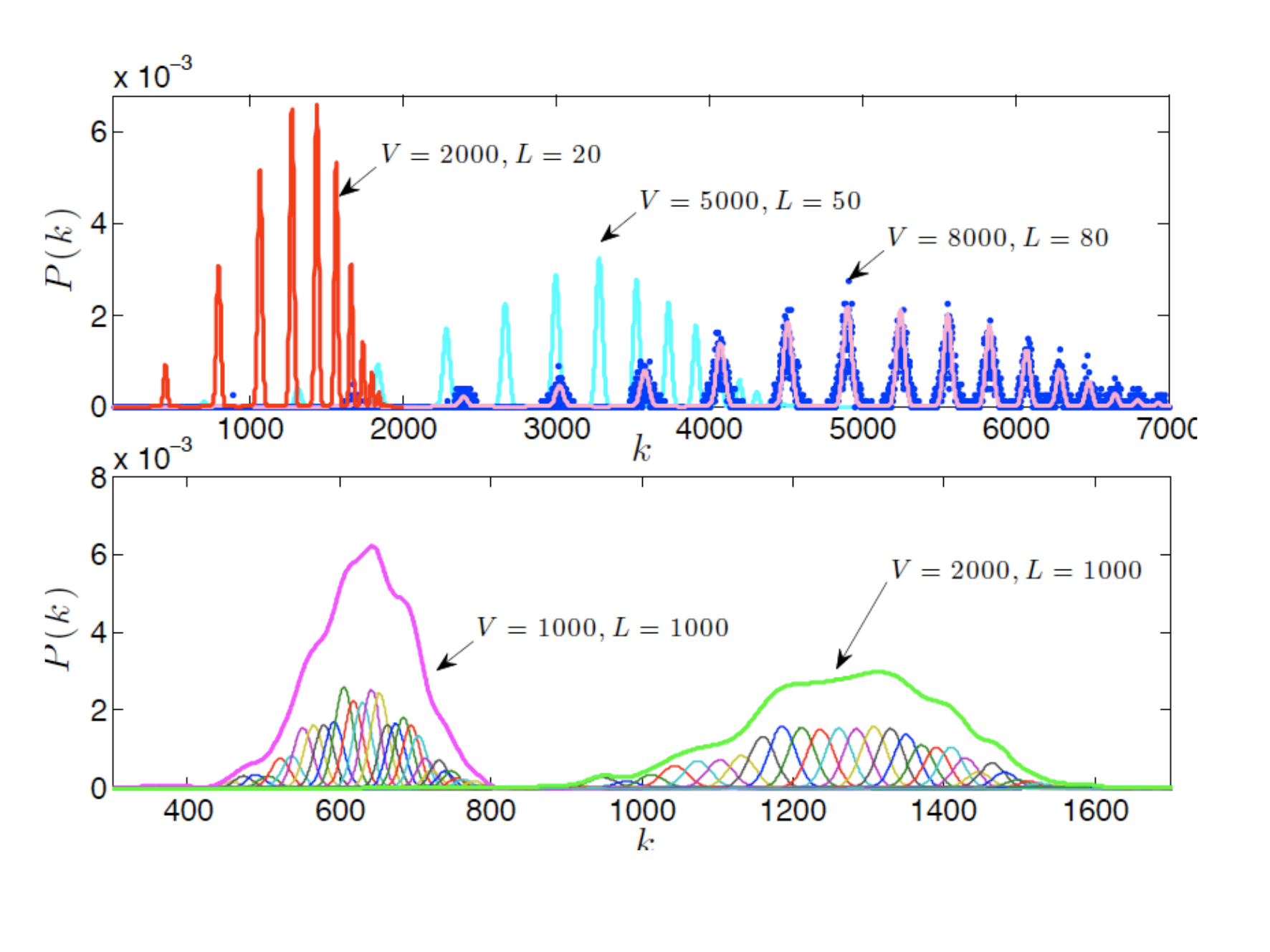}
\caption{\label{fig:distri} Degree distribution
$\bar{P}_{\mathrm{degree}}(k)$ for systems displaying small values
of $L$ and multimodal distribution (upper panel) and large values
of $L$ and distribution collapsing into a unimodal one (lower
panel). In the former case we compare systems of different sizes
but fixed $\alpha = 0.01$, where  continuous lines represent the
analytic estimate of eq. (\ref{eq:overall_degree}) while symbols
($\bullet$) represent data from simulations and are reported for
clearness only for the case $V=8000$. In the lower panel we
compare systems with same $L$ but different volumes; thicker
curves represent $\bar{P}_{\mathrm{degree}}(z;a,L,V)$, while
each mode $P_{\mathrm{degree}}(z;\rho,a,V)$ is depicted in
different colors.
}
\end{figure}

Now, the overall distribution can be written as a combination of binomial distributions
\begin{equation} \label{eq:overall_degree}
\bar{P}_{\mathrm{degree}}(z; a,L, V) = \sum_{\rho=0}^{L}
P_{\mathrm{degree}}(z; \rho, a, V)P_1(\rho;a,L),
\end{equation}
where the overlap among two ``modes'', say $\rho$ and $\rho+1$, can
be estimated through $\sigma_{\rho}/(\bar{z}_{\rho +1 } -
\bar{z}_{\rho})$: Exploiting eqs. (\ref{eq:average_degree_rho})
and (\ref{eq:variance_rho}) we get
\begin{equation} \label{eq:multi}
\frac{\sigma_{\rho}}{\bar{z}_{\rho + 1} - \bar{z}_{\rho}} =
\sqrt{1 - \left( \frac{1-a}{2} \right)^{\rho} } \left[ \sqrt{V}
\left( \frac{1-a}{2} \right)^{\rho/2} \left( \frac{1+a}{2} \right)
\right]^{-1} \sim \sqrt{\frac{L}{V}} = \sqrt{\alpha},
\end{equation}
where the generic mode $\rho$ is confused with $\bar{\rho}$ and
the approximate result $\sqrt{L/V}$ was derived by using the
scaling $a=-1+\gamma/V^{\theta}$, with $1/2 < \theta \leq 1$
(both these points are fully discussed in the next Section); also,
the last passage holds rigorously in the thermodynamic limit of
the  high storage regime ($L$ linearly diverging with $V$).
Interestingly,
for systems with different scaling regimes among $L$ and $V$, for instance $L \propto \log V$ \cite{baglia,agliba}, the distribution remains multi-modal because a vanishing
overlap occurs among the single distributions
$P_{\mathrm{degree}}(z; \rho, a, V)$: $\bar{P}_{\mathrm{degree}}(z;
 a, L, V)$ turns out to be an $(L+1)$-modal distribution (see
Fig.~{\ref{fig:distri}, upper panel); vice versa, for $L \propto
V$, the overall distribution gets
mono-modal
(see Fig.~{\ref{fig:distri}, lower panel). Briefly, we mention that for $\theta=1/2$ the ratio in the l.h.s. of eq.~(\ref{eq:multi}) still converges to a finite value approaching $\sqrt{\alpha}$ for $\gamma^2 \ll \alpha$, while for $\theta <1/2$ it diverges.

From eq. (\ref{eq:overall_degree}), the average degree for a
generic node  is
\begin{equation}\label{eq:average_degree_rho2}
\bar{z} = \sum_{z=0}^V z \, \bar{P}_{\mathrm{degree}}(z;  a, L,V) = \sum_{\rho=0}^L P_1(\rho; a;L) \bar{z}_{\rho} = V \left\{ 1 - \left[ 1 -  \left( \frac{1+a}{2}  \right)^{2} \right]^L \right\},
\end{equation}
where
\begin{equation} \label{eq:p_link_av}
p = 1 - \left[ 1 -  \left( \frac{1+a}{2}  \right)^{2} \right]^L
\end{equation}
is  the average link probability for two arbitrary strings $\xi_i$
and $\xi_j$, which can be obtained by averaging over all possible
string arrangements, namely, recalling eqs. (\ref{eq:rho}) and
(\ref{eq:prob2}),
\begin{eqnarray}
\nonumber
p &=&
\sum_{\rho_i=0}^L \sum_{\rho_j=0}^L P_1(\rho_i; a, L) P_1(\rho_j; a, L) P_{\mathrm{link}}(\rho_i,\rho_j; a, L) \\
\nonumber
&=& 1 - \left ( \frac{1-a}{2} \right)^{2L} \sum_{\rho_i=0}^L \sum_{\rho_j=0}^L \left ( \frac{1+a}{1-a} \right)^{\rho_i+\rho_j} \binom{L}{\rho_i} \binom{L-\rho_i}{\rho_j}\\
\label{eq:p_easy}
&=& 1 - \left[ 1- \left( \frac{1+a}{2}\right)^2 \right]^L.
\end{eqnarray}
Of course, eq. (\ref{eq:p_easy}) could be obtained directly by
noticing that the probability for the $\mu$-th entries of two
strings not to yield any contribute is $1 - [(1+a)/2]^2$, so that
two strings are connected if there is at least one matching.


\subsection{Coupling distribution}\label{sec:Couplings}
As explained in Sec.~\ref{sec:model}, the coupling $J_{ij}$ among
nodes $i$ and $j$ is given by the relative number of matching
entries among the corresponding strings $\xi_i$ and $\xi_j$. Eq.~
 (\ref{eq:fond}) provides the probability for $\xi_i$ and $\xi_j$ to
share a link of magnitude $J=k$, namely $P_{\mathrm{coupling}}(J;\rho_i,\rho_j,L) = P_{\mathrm{match}}(k;\rho_i,\rho_j,L) $.
Following the same arguments as in the previous section we get the probability that a link stemming from $\xi_i$ has magnitude $J$, that is
\begin{equation} \label{eq:coupling_distr}
\bar{P}_{\mathrm{coupling}}(J; \rho_i,a) = \sum_{\rho_j=0}^L P_{\mathrm{coupling}}(J;  \rho_i,\rho_j, L) P_1(\rho_j; a, L) =   \binom{\rho_i}{J}  \left( \frac{1-a}{2} \right) ^{\rho_i-J} \left( \frac{1+a}{2} \right) ^{J},
\end{equation}
which is just the probability that $J$ out of $\rho_i$ non-null entries are properly matched with the generic second node.


Similarly to  $\bar{P}_{\mathrm{degree}}(z; a, L, V)$, the overall coupling
distribution can be
written as the superposition
$ \sum_{\rho} P_1(\rho; a,L)
\bar{P}_{\mathrm{coupling}}(J; \rho,a)$, giving rise to a multimodal
distribution. Each mode has variance $\sigma_{\rho}^2 = \rho (1- a^2) /4$ and is peaked at
\begin{equation}\label{eq:average_J_rho}
\bar{J}_{\rho} = \rho \; \frac{1+a}{2},
\end{equation}
which represents the average coupling expected for  links stemming
from a node with $\rho$ non-null entries. Nevertheless, by comparing $\bar{J}_{\rho+1} - \bar{J}_{\rho} =
(1+a)/ 2$ and the standard deviation $\sqrt{\rho(1-a^2)}/2$, we
find that in the limit $L=\alpha V$ and $V \rightarrow \infty$ the distribution
gets mono-modal.

Anyhow, we can still define the average weighted degree $w_{\rho}$ expected for a node displaying $\rho$ non-null entries. Given that for the generic node $i$, $w \equiv \sum_{j} J_{ij}$, we get
\begin{equation} \label{eq:average_weightdegree}
\bar{w}_{\rho} = V \bar{J}_{\rho} = V \; \rho   \; \frac{1+a}{2}.
\end{equation}
Of course, one expects that the larger the coordination
number of a node and the larger its weighted degree; such a
correlation is linear only in the regime of low connectivity. In
fact, by merging eq. (\ref{eq:average_degree_rho}) and eq.
(\ref{eq:average_weightdegree}), one gets
\begin{equation} \label{eq:w_k}
\bar{w}_{\rho} = \left( \frac{1+a}{2} \right) \frac{\log \left( 1 - \frac{ \bar{z}_{\rho}}{V} \right)}{\log \left(\frac{ 1-a}{2} \right)} V \approx  \bar{z}_{\rho},
\end{equation}
where the last expression holds for $ \bar{z}_{\rho} \ll V$ and $a \ll 1$.

It is important to stress that (apart pathological cases which
will be taken into account in the $L\to \infty$ scaling later) the
variance of $\rho$ scales as $\sigma_{\rho}^2(a;L) = (1-a^2)L/4$
such that, despite the average of $\rho$ is $(1+a)L/2$,
substituting $\rho/L$ with $(1+a)/2$ into eq. $(3.18)$ becomes
meaningless in the thermodynamic limit as the variance of
$\bar{J}_{\rho}$ diverges as $\sqrt{L} \propto \sqrt{V}$: This
will affect drastically the thermodynamics whenever far from the
Curie-Weiss limit.

It should be remarked that $\bar{J}_{\rho}$ represents the average coupling for a link stemming from a node characterized by a string with $\rho$ non-null entries, where the average includes also non-existing links corresponding to zero coupling. On the other hand, the ratio $\bar{w}_{\rho}/\bar{z}_{\rho}$ directly provides the average magnitude for existing couplings.
Moreover, the average magnitude for a generic link is
\begin{equation} \label{eq:average_J}
\bar{J} = \sum_{\rho=0}^L P_1(\rho; a, L) \bar{J}_{\rho} = \left( \frac{1+a}{2} \right)^2.
\end{equation}

By comparing eq. (\ref{eq:average_J_rho}) and eq.
(\ref{eq:average_J}) we notice that the local energetic
environment seen by a single node, i.e. $\bar{J}_{\rho}$, and the
overall energetic environment, i.e. $\bar{J}$, scale,
respectively, linearly and quadratically with $(1+a)/2$: we will
see in the thermodynamic dedicated section that (apart in the
Curie-Weiss limit where global and local effects merge) despite
the self-consistence relation (which is more sensible by local
condition) will be influenced by $\sqrt{\bar{J}}$, critical
behavior will be found at $\beta_c = \bar{J}^{-1}$ coherently with
a manifestation of a collective, global effect.

Anyhow, when $V$ is large and the coupling distribution is
narrowly peaked at the mode corresponding to $\bar{\rho}_{a,L}$,
the couplings can be rather well approximated by the average value
$\bar{J}_{L(1+a)/2} = [(1+a)/2]^2 = \bar{J}$, so that the disorder
due to the weight distribution may be lost; as we will show this
can occur in the regime of high dilution ($\theta > 1/2$). As for
the other source of disorder (i.e. topological inhomogeneity),
this can also be lost if $a$ is sufficiently larger than $-1$ as
we are going to show.


\subsection{Scalings in the thermodynamic limit} \label{ssec:scaling}
In the thermodynamic limit and high-storage regime,  $L$ is
linearly divergent with $V$ and the average probability $p$ for two nodes
to be connected (see eq. (\ref{eq:p_link_av})) approaches a
discontinuous function assuming value $1$ when $a>-1$, and value
$0$ when $a=-1$. More precisely, as $V \rightarrow \infty$ there
exists a vanishingly small range of values for $a$ giving rise to a
non-trivial graph; such a range is here recognized by the following scaling
\begin{equation}\label{eq:scaling}
a = -1 + \frac{\gamma}{V^{\theta}},
\end{equation}
where $\theta \geq 0$ and $\gamma$ is a finite parameter.

First of all, we notice that, following eqs.~(\ref{eq:rho_av}) and (\ref{eq:rho_var}),
\begin{eqnarray}
\bar{\rho}_{-1+\gamma/V^{\theta}, \alpha V} =\frac{\alpha \gamma}{2 V^{\theta -1}} \\
\sigma_{-1+\gamma/V^{\theta}, \alpha V}^2 = \frac{\alpha \gamma}{2 V^{\theta -1}} \left( 1- \frac{\gamma}{2 V^{\theta}} \right) \sim \bar{\rho}_{-1+\gamma/V^{\theta}, \alpha V},
\end{eqnarray}
where the last approximation holds in the thermodynamic limit and it is consistent with the convergence of the binomial distribution in eq.~(\ref{eq:rho}) to a Poissonian distribution. For $\theta \leq 1$, $\bar{\rho} \gtrsim  \sigma$, so that when referring to a generic mode $\rho$, we can take without loss of generality $\bar{\rho}$; the case $\theta >1$ will be neglected as it corresponds to a disconnected graph.

Indeed, the probability for two arbitrary nodes to be connected gets
\begin{equation} \label{eq:giusto}
p = 1 - \left[ 1 - \left( \frac{1+a}{2} \right)^2 \right]^L =  1 - \left[ 1 - \frac{\gamma^2}{4V^{2 \theta}} \right]^{\alpha V}  \underset{V
\to \infty}{\to}  1 - e^{-\gamma^2 \alpha V^{1- 2 \theta} /4},
\end{equation}
so that we can distinguish the following regimes:
\begin{itemize}
\item  $\theta <1/2$, \; $p \approx 1$, \; $\bar{z} \approx V$  \; $\Rightarrow$  Fully connected (FC) graph\\
\item  $\theta =1/2$, \; $p \sim 1- e^{-\gamma^2 \alpha/4} \sim \gamma^2 \alpha/4$, \;  $\bar{z} = O(V)$ \; $\Rightarrow$ Linearly diverging connectivity \\
Within a mean-field description the Erd\"{o}s-R\'{e}nyi (ER) random graph with finite probability $\mathcal{G}(V,p)$ is recovered.
\item  $1/2 < \theta <1$, \; $p \sim \gamma^2  \alpha V^{1-2 \theta}/4$, \; $\bar{z} = O(V^{2 - 2 \theta})$ \; $\Rightarrow$  Extreme dilution regime (ED)\\
In agreement with \cite{watkin,tonED}, $\lim_{V \rightarrow \infty}  \bar{z}^{-1} = \lim_{V \rightarrow \infty} \bar{z}/V = 0$.\\
\item  $ \theta = 1$, \; $p \sim \frac{\gamma^2  \alpha}{4V}$, \; $\bar{z} = O(V^{0})$ \; $\Rightarrow$  Finite connectivity regime\\
Within a mean-field description $\gamma^2 \alpha / 4 = 1$ corresponds to a percolation threshold.
\end{itemize}

Therefore, while $\theta$ controls the connectivity regime of the network, $\gamma$ allows a fine tuning.

As for the average coupling (see eq. (\ref{eq:average_J})) and the average weighted degree:
\begin{equation}\label{eq:JJ}
\bar{J} =   \frac{\gamma^2}{4 V^{2 \theta}},
\end{equation}
\begin{equation}
\bar{w} =  V \bar{J} =  \frac{\gamma^2}{4 V^{2 \theta-1}}.
\end{equation}
Now, the average ``effective coupling'' $\tilde{J}$, obtained by averaging only on existing links, can be estimated as
\begin{equation} \label{eq:cases}
 \tilde{J} =  \bar{J} /  p  = \left\{ \begin{array}{rl}
 \gamma^2/(4 V^{2 \theta})  &\mbox{ if $\theta < 1/2$} \\
\gamma^2 / [4 V^{2 \theta} (1 - e^{-\gamma^2 \alpha/4}) ] &\mbox{ if $\theta = 1/2$}\\
  1/(\alpha V) = 1/L &\mbox{ if $1/2 < \theta \leq 1$}
       \end{array} \right.
\end{equation}
Interestingly, this results suggests that in the thermodynamic limit, for values of $a$ determined by eq.~(\ref{eq:scaling}) with $1/2 < \theta \leq 1$, nodes are pairwise either non-connected or connected due to one single matching among the relevant strings. This can be shown more rigorously by recalling the coupling distributions $\bar{P}_{\mathrm{coupling}}(J; \rho_i, L)$ of eq.~(\ref{eq:coupling_distr}): In particular, for $\theta >1/2$, neglecting higher order corrections, for $J=0$ the probability is $p_0 \sim \exp(\alpha \gamma^2 V^{1 - 2 \theta}/4) \sim 1- \gamma^2 \alpha / (4 V^{2 \theta -1})$, for $J=1/L$ the probability is $p_1 \sim p_0 \gamma^2 \alpha / (4 V^{2 \theta -1}) \sim 1 - p_0 $. For $\theta =1/2$ this still holds for $\alpha \gamma^2 /4 \ll 1$, which corresponds to a relatively high dilution regime, otherwise some degree of disorder is maintained, being that $p_k \sim (\alpha \gamma^2/4)^k / k!$. On the other hand, for $\theta < 1/2$, while topological disorder is lost (FC), the disorder due to the coupling distribution is still present. However, notice that for $\theta = 0$ and $\gamma = 2$, $\bar{P}_{\mathrm{coupling}}(J; \rho_i, L)$ gets peaked at $J=L$ and, again, disorder on couplings is lost so that a pure Curie-Weiss model is recovered.

This means that, for $L=\alpha V$ and $V \to \infty$, we can
distinguish three main regions in the parameter space $(\theta,
\alpha, \gamma)$ where the graph presents only topological
disorder ($\theta >1/2$), or only coupling disorder ($\theta
<1/2$), or both ($\theta =1/2 \wedge \gamma^2 \alpha = O(1)$).

In general, we expect that the the critical temperature scales
like the connectivity times the average coupling and the system
can be looked at as a fully connected with average coupling equal
to $\bar{J}$ or as a diluted network with effective coupling
$\tilde{J}$ and connectivity given by $\bar{z}$; in any case we
get $\beta^{-1}_c \sim \bar{J}$ (crf. eq.$(4.37)$).


\subsection{Small-world properties} \label{ssec:SW}
Small-world networks are endowed, by definition, with high cluster coefficient, i.e. they display sub-networks that are characterized by the presence of connections between almost any two nodes within them, and with small diameter, i.e. the mean-shortest path length among two nodes grows logarithmically (or even slower) with $V$. While the latter requirement is a common property of random graphs \cite{newman,albert},
the clustering coefficient deserves much more attention also due to the basic role it covers in biological \cite{pheno,herpes} and social networks \cite{grano1,grano2}.

The clustering coefficient measures the likelihood that two neighbors of a node are linked themselves; a higher clustering coefficient indicates a greater ``cliquishness''.
Two versions of this measure exist \cite{newman,albert}: global
and local; as for the latter the coefficient $c_i$ associated to a
node $i$ tells how well connected the neighborhood of $i$ is. If
the neighborhood is fully connected, $c_i$ is
$1$, while a value close to $0$ means that there are hardly any
connections in the neighborhood.

The clustering coefficient of a node is defined as the ratio between the number of
connections in the neighborhood of that node and the number of
connections if the neighborhood was fully connected. Here
neighborhood of node $i$ means the nodes that are connected to $i$
but does not include $i$ itself. Therefore we have
\begin{equation}
c_i = \frac{2 E_i}{z_i (z_i -1)},
\end{equation}
where $E_i$ is the number of actual links present, while
$z_i(z_i-1)/2$ is the number of connections for a fully connected
group of $z_i$ nodes. Of course, for the Erd\"{o}s-Renyi graph
where each link is independently drawn with a probability $p$, one
has $c^{ER} = p$, regardless of the node considered.

We now estimate the clustering coefficient for the graph
$\mathcal{G}(a,L,V)$, focusing the attention on a range of $a$
such that the average number of non-null entries per string is
small enough for the link probability to be strictly lower than
$1$ so that the topology is non trivial; to fix ideas and
recalling last section $1/2 \leq \theta \leq 1$. Let us consider a
string displaying $\rho$ non-null entries, corresponding to the
positions $\mu_1, \mu_2, ..., \mu_{\rho}$, and $z$
nearest-neighbors; the latter can be divided in $\rho$ groups:
strings belonging to the $j$-th group have $\xi^{\mu_j}=1$.
Neglecting the possibility that a nearest-neighbor can belong to
more than one group contemporary (in the thermodynamic limit this
is consistent with Eq.~\ref{eq:cases}), we denote with $n_j$ the
number of nodes belonging to the $j$-th group, being $\sum_{j} n_j
= z$, whose average value is $z/\rho$ (which, due to the above
assumptions is larger than one). Now, nodes belonging to the same
group are all connected with each other as they share at least one
common trait, i.e. they form a clique; the contribute of
intra-group links is
\begin{equation} \label{eq:intra}
E_{\mathrm{intra}} = \frac{1}{2} \sum_{i=1}^{\rho} n_i(n_i -1) =  \frac{1}{2} \left( \sum_{i=1}^{\rho} n_i^2 - z \right) \approx   \frac{1}{2}  \left[ \left( \frac{z}{\rho}\right)^2 \rho - z \right],
\end{equation}
while the contribute of inter-group links can be estimated as
\begin{equation} \label{eq:inter}
E_{\mathrm{inter}} \approx \sum_{i,j=1, i\neq j}^{\rho} n_i n_j \tilde{p} \approx \left( \frac{z}{\rho}\right)^2  \binom{\rho}{2} \, \tilde{p},
\end{equation}
where $\tilde{p}$ is the probability for two nodes linked to $i$
and belonging to different groups to be connected, and the sum
runs over all possible $\binom{\rho}{2}$ couples of groups. Hence, the total
number of links among neighbors is $E=E_{\mathrm{intra}} +
E_{\mathrm{inter}} = \{ \sum_{i=1}^{\rho} n_i \sum_{j=1}^{\rho}
n_j [\tilde{p} +(1-\tilde{p})\delta_{ij}] - z \} /2$, where
$\delta_{ij}$ is the Kronecker delta returning $1$ if $i=j$ and
zero otherwise; of course, for $\tilde{p}=1$ we have $E=(z^2 - z)/2$ and
$c_i=1$.

Now, in the average, the probability $\tilde{p}$ is smaller than
$p$ as it represents the probability for two strings of length
$L-1$ and displaying an average number of non-null entries equal
to $\rho -1$ to be connected. However, for $\rho$ and $L$ not too
small the two probabilities converge so that by summing the two
contributes in eq. (\ref{eq:intra}) and (\ref{eq:inter}) we get
\begin{equation}
E \approx \frac{1}{2}  \left[ \left( \frac{z}{\rho}\right)^2 \rho
- z \right] + \left( \frac{z}{\rho}\right)^2  \binom{\rho}{2} \,
\tilde{p} \Rightarrow c \approx p + \frac{1}{\rho} - \frac{1}{z-1} > p,
\end{equation}
where in the last inequality we used $\rho < z -1$.
Therefore, it follows straightforward that $c_i$ is larger than
the clustering coefficient expected for an ER graph displaying the
same connectivity, that is $c^{ER} = p$.

\begin{figure}[tb]\bc
\includegraphics[height=100mm]{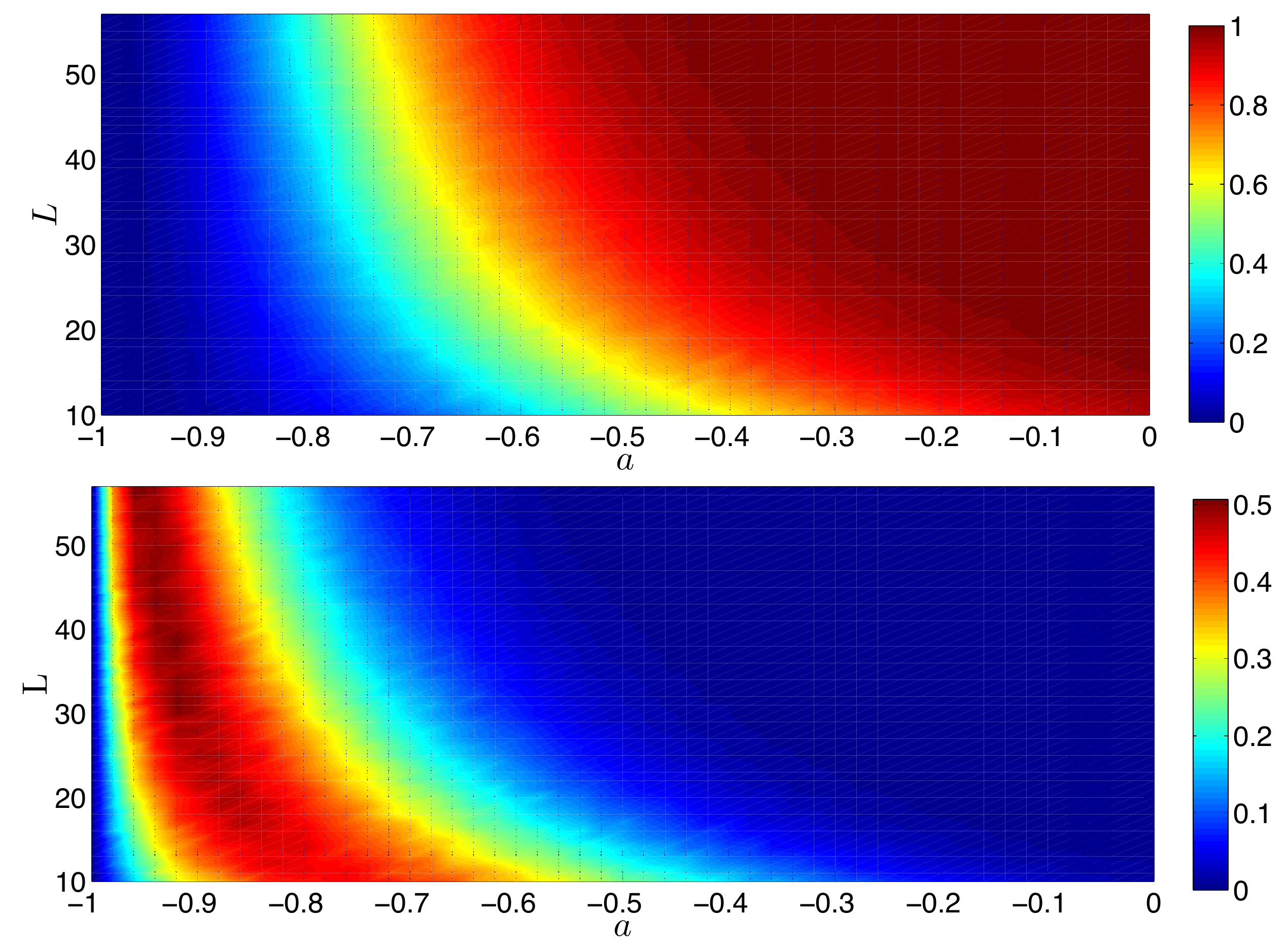}
\caption{\label{fig:clustering} Upper panel: average link probability $p = \bar{z}/N$; Lower panel: difference between the average clustering coefficient for $\mathcal{G}(V,L,a)$ and for an analogous ER graph just corresponding to $p$. Both plots are presented as function of $a$ and $L$ and refer to a system of $V=2000$ nodes.} \ec
\end{figure}

From previous arguments  it is clear that the SW effect gets more
evident, with respect to the ER case taken as reference, when the
network is highly diluted. This is confirmed by numerical data:
Fig.~\ref{fig:clustering} shows in the lower panel the clustering
coefficient expected for the analogous ER graph, namely
$c^{ER}=\bar{z} /V$, while in the upper panel it shows the
difference between the average local clustering coefficient $c =
\sum_{i=1}^V c_i/V$ and $c^{ER}$ itself. Of course, when $a$
approaches $1$, the graph gets fully connected and $c \to c^{ER} \to 1$.

Finally we mention that when focusing on the low storage regime, a non-trivial distribution for couplings can give rise to interesting effects. Indeed, weak ties can be shown \cite{forth} to work as bridges connecting communities strongly  linked up, as typical of real networks  \cite{grano1,neuro}.
Also, as often found in technological and biological
networks, the graph under study display a ``dissortative mixing'' \cite{newman,albert}, that is to say, high-degree
vertices prefer to attach to low-degree nodes \cite{forth}.

\section{Thermodynamics}

So far the emergent network has been exhaustively  described
by a random, correlated graph  whose links are endowed with
weights; we now build up a
quantitative thermodynamics on such a structure.

Once the Hamiltonian $H_V(\sigma;\xi)$ is given (eq.~
\ref{modello}), we can introduce the partition function
$Z_V(\beta;\xi)$ as \be Z_V(\beta;\xi) = \sum_{\sigma}e^{-\beta
H_V(\sigma; \xi),} \ee the Boltzmann state $\omega$ as \be
\omega(.)=\frac{\sum_{\sigma}. e^{-\beta
H_V(\sigma;\xi)}}{Z_V(\beta;\xi),} \ee and the related free energy
as \be A(\beta,\alpha, a) = \lim_{V\to\infty}\frac1V \mathbb{E}
\log Z_V(\beta;\xi),\ee where $\mathbb{E}$ averages over the
quenched distributions of the affinities $\xi$.
\newline
Once the free energy (or equivalently the {\em pressure}) is
obtained, remembering that (calling $S$ the entropy and $U$ the
internal energy)
$$
A(\beta,\alpha,a)=-\beta f(\beta,\alpha,a) = S(\beta,\alpha,a) -
\beta U(\beta,\alpha,a),$$ the whole macroscopic properties,
thermodynamics, can be derived due the Legendre structure of
thermodynamic potentials \cite{whiteman}.

\subsection{Free energy trough extended double stochastic
stability}\label{double}
%
For the sake of clearness now
we expose in complete generality and details the whole plan
dealing with a generic expectation on $\xi$ (i.e. $\mathbb{E}\xi =
(1+a)/2$), then, we will study the $L\to\infty$ scaling, in which
$a$ must tend to $-1$ more carefully.
\newline
With this palimpsest in mind, let us normalize the Hamiltonian
$(2.2)$ in a more convenient form for this section (i.e. dividing
by $L$ the $J_{ij}$, such that the effective coupling is bounded
by $1$), and let us neglect the external field $h$ which can be
implemented later straightforwardly.
\begin{equation}
H_V(\sigma;\xi)=
\frac{1}{VL}\sum_{ij}^V\sum_{\mu}^L \xi_i^{\mu}\xi_j^{\mu}\sigma_i
\sigma_j.
\end{equation}
As a next step, through the Hubbard-Stratonovick transformation
\cite{whiteman,ellis}, we map the partition function of our
Hamiltonian into a bipartite Erd\"{o}s-R\'{e}nyi ferromagnetic
random graph \cite{ABC}\cite{genovese}, whose parties are the
former built by the $V$ agents and a new one built of by $L$
Gaussian variables $z_{\mu}$, $\mu \in (1,...,L)$: \be
Z(\beta;\xi) = \sum_{\sigma}\exp\big(-\beta H_{V}(\sigma;\xi)\big)
= \sum_{\sigma}\int_{-\infty}^{+\infty}\prod_{\mu=1}^L
d\mu(z_{\mu}) \exp\Big(\sqrt{\frac{\beta}{L V}} \sum_{i}^{V}
\sum_{\mu}^L \xi_{i,\mu}\sigma_i z_{\mu} \Big), \ee where with
$\prod_{\mu=1}^L d\mu(z_{\mu})$  we mean the Gaussian measure on
the product space of the Gaussian party. Note that, even when $L$
goes to infinity linearly with $V$ (as in the high storage
Hopfield model \cite{AGS1}), due to the normalization encoded into
the affinity product of the $\xi$'s nor the $z$-diagonal term
contribute to the free energy (as happens in the neural network
counterpart \cite{BGG}), neither (but this will be clear at the
end of the section) there is a true dependence by $\alpha$ in the
thermodynamics.
\newline
Furthermore, notice that the graph of the interactions among the two
parties is now a simple, and no longer weighted, Erd\"{o}s-R\'{e}nyi
\cite{barabasi}: so we started with a complex topology for a
single party and we turned this problem in solving the
thermodynamics for a simpler topology but paying the price of
accounting for another party in interaction. The lack of weight
on links will have fundamental importance when defining the
order parameters.
\newline
Another approach to this is noticing that if we dilute -randomly-
directly the Hopfield model (i.e. as checking for its robustness
as already tested by Amit \cite{amit}) we push it on an
Erdos-Renyi topology, while if we dilute its entries in pattern
definitions (due to the Hebbian kernel) we have to deal with
correlated dilution.
\newline
Consequently (strictly speaking assuming the existence of the $V$
limit) we want to solve for the following free energy: \be
A(\beta,\alpha,a) = \lim_{V\to\infty}
\frac1V\mathbb{E}\log\sum_{\sigma}\int_{-\infty}^{+\infty}
\prod_{\mu}^L d\mu(z_{\mu}) \exp\Big( \sqrt{\frac{\beta}{L V}}
\sum_{i}^{V} \sum_{\mu}^L \xi_{i,\mu}\sigma_i z_{\mu} \Big). \ee
To this task we extend the method of the double stochastic
stability recently developed in \cite{BGG} in the context of
neural networks. Namely we introduce independent random fields
$\eta_i, i \in (1,...,V)$ and $\chi_{\mu}, \mu \in (1,...,L)$,
(whose probability distribution is the same as for the $\xi$
variables -as in every cavity approach-), which account for
one-body interactions for the agents of the two parties. So our
task is to interpolate among the original system and the one left
with only these random perturbations: Let us use $t\in[0,1]$ for
such an interpolation; the trial free energy $A(t)$ is then
introduced as follows
\begin{eqnarray}\label{interpolante} A(t) &=& \lim_{V\to\infty} \frac1V \mathbb{E} \log
\sum_{\sigma}\int_{-\infty}^{+\infty} \prod_{\mu}^L d\mu(z_{\mu}) \cdot \\
\nonumber &\cdot& \exp\Big( t \sqrt{\frac{\beta}{L V}}\sum_{i
\mu}^{V L} \xi_{i \mu} \sigma_i z_{\mu} + (1-t) [ \sum_{l_c=1}^L
b_{l_c} \sum_i^V \eta_i \sigma_i + \sum_{l_b=1}^V c_{l_b}
\sum_{\mu}^L \chi_{\mu} z_{\mu} \Big),
\end{eqnarray} where now
$\mathbb{E}=\mathbb{E}_{\xi}\mathbb{E}_{\eta}\mathbb{E}_{\chi}$
and $b_{l_c}$ [with $l_c \in(1,...,L)$], and  $c_{l_b}$ [with $l_b
\in(1,...,V)$]  are real numbers (possibly functions of
$\beta,\alpha$) to be set a posteriori.
\newline
As the theory is no longer Gaussian, we need infinite sets of
random fields (mapping the presence of multi-overlaps in standard
dilution\cite{ABC}\cite{gds} and no longer only the first two
momenta of the distributions).
\newline
Of course we recover the proper free energy by evaluating the
trial $A(t)$ at $t=1$, $(A(\beta, \alpha, a)=A(t=1))$, which we
want to obtain by using the fundamental theorem of calculus: \be
A(1) = A(0) + \int_0^1 \Big( \partial A (t') / \partial t'
\Big)_{t'=t}dt. \ee To this task we need two objects: The trial
free energy $A(t)$ evaluated at $t=0$ and its $t$-streaming
$\partial_t A(t)$.
\newline
Before outlining the calculations, some definitions are
in order here to lighten the notation: taken $g$ as a generic
function of the quenched variables we have
\begin{eqnarray}
&&\mathbb{E}_{\eta} g(\eta) = \sum_{l_b=0}^{V} P(l_b)
g(\eta_{l_b}) = \sum_{l_b=0}^{V}  \binom{V}{l_b}
\left(\frac{1+a}{2}\right)^{V-l_b}\left(\frac{1-a}{2} \right)^{l_b} g(\eta_{l_b}), \\
&& \mathbb{E}_{\chi} g(\chi) = \sum_{l_c=0}^{L} P(l_c)
g(\chi_{l_c}) =
  \sum_{l_c=0}^{L}  \binom{L}{l_c} \left(\frac{1+a}{2} \right)^{L-l_c}\left(\frac{1-a}{2}\right)^{l_c} g(\chi_{l_c}), \\
&& \mathbb{E}_{\xi}g(\xi) =
\sum_{l_b=0}^{V}\sum_{l_c=0}^{L}\binom{V}{l_b}\binom{L}{l_c}
\left(\frac{1+a}{2}\right)^{l_b+l_c}\left(\frac{1-a}{2}\right)^{V+L-l_b-l_c}\delta_{l_bl_c=l},
\end{eqnarray}
where $P(l_b)$ is the probability that $l_b$ (out of $V$ random
fields) are active, i.e. $\eta=1$, so that the number of spins
effectively contributing to the function $g$ is $l_b$;
analogously, mutatis mutandis, for $P(l_c)$. Moreover, in the last
equation we summed over the probability $P(l)$ that in the
bipartite graph a number $l$ of links out of the possible $V
\times L$ display a non-null coupling, i.e. $ \xi \neq 0 $;
interestingly, eq. (4.10) can be rewritten in terms of the
above mentioned $P(l_b)$ and $P(l_c)$. In fact, $\xi_{i,\mu}$ can
be looked at as an $V \times L$ matrix generated by the product of
two given vectors like $\eta$ and $\chi$, namely
$\xi_{i,\mu}=\eta_i \chi_{\mu}$, in such a way that the number of
non-null entries in the overall matrix $\mathbf{\xi}$ is just
given by the number of non-null entries displayed by $\eta$ times
the number of non-null entries displayed by $\chi$. Hence, $P(l)$
is the product of $P(l_b)$ and $P(l_c)$ conditional to $l_b l_c
=l$. 

\subsection{The `topologically microcanonical" order parameters}

 Starting with the streaming of eq. (\ref{interpolante}),
 this operation gives raise to the sum of
three terms $\mathcal{A}+\mathcal{B}+\mathcal{C}$. The former when
deriving the first contribution into the exponential, the last two
terms when deriving the two contributions by all the $\eta$ and
$\chi$.
\begin{eqnarray}
\mathcal{A} &=& +\frac1V
\sqrt{\frac{\beta}{LV}}\sum_{i,\mu}^{V,L}\mathbb{E} \xi_{i \mu}
\omega (\sigma_i z_{\mu}) = \sqrt{\alpha\beta} \left(\frac{1+a}{2}\right) \sum_{l_b,l_c}^{V,L} P(l_b)P(l_c) M_{l_b}N_{l_c}\\
\mathcal{B} &=& -\sum_{l_c=1}^L \frac{b_{l_c}}{V} \sum_i^V
\mathbb{E}\eta_i
\omega(\sigma_i) = - \sum_{l_c=1}^L b_{l_c} \left(\frac{1+a}{2}\right) \sum_{l_b=0}^{V}P(l_b) M_{l_b} \\
\mathcal{C} &=& -\sum_{l_b=1}^V \frac{c_{l_b}}{N} \sum_{\mu}^L
\mathbb{E} \chi_{\mu} \omega(z_{\mu}) = -
\sqrt{\alpha}\sum_{l_b=1}^V c_{l_b} \left(\frac{1+a}{2}\right)
\sum_{l_c=0}^{L} P(l_c) N_{l_c},
\end{eqnarray}
where we introduced the following order parameters
\begin{eqnarray}
M_{l_b}= \frac1V \sum_i^V \omega_{l_b+1}(\sigma_i), \\
N_{l_c}= \frac1L \sum_{\mu}^L \omega_{l_c+1}(z_{\mu}),
\end{eqnarray}
and the Boltzmann states $\omega_k$ are defined by taking into
account only $k$ terms among the elements of the party involved.
\newline
Of course the Boltzmann states are no longer the ones introduced
into the definition (4.2) but the extended ones taking into
account the interpolating structure of the cavity fields (which
however will recover the originals of statistical mechanics when
evaluated at $t=1$).
\newline
Namely, $\omega_{l_b+1}$ has only $l_b+1$ terms of the type $b
\sigma$ in the Maxwell-Boltzmann exponential, ultimately
accounting for the (all equivalent in distribution) $l_b+1$ values
of $\eta=1$, all the others being zero.
\newline
In the same way $\omega_{l_c+1}$ has only $l_c+1$ terms of the
type $c z$ in the Maxwell-Boltzmann exponential, ultimately
accounting for the (all equivalent in distribution) $l_c+1$ values
of $\chi=1$, all the others being zero.
\newline
When dealing with $\xi_{i \mu}$ we can decompose the latter
accordingly to what discussed before.  By these ``partial Boltzmann
states" we can define the averages of the order parameters as
\begin{eqnarray}
\langle M \rangle &=&
\sum_{l_b}^{V-1}P(l_b)M_{l_b},
\\
\langle N \rangle &=& \sum_{l_c}^{L-1}P(l_c)N_{l_c}.
\end{eqnarray}
These objects  may deserve more explanations because, as a main
difference with classical approaches
\cite{ABC}\cite{monta}\cite{gds}, here replicas and their overlaps
are not involved (somehow suggesting the implicit correctness of a
replica symmetric scenario). Conversely, we do conceptually two
(standard) operations when introducing our order parameters: at
first we average over the ($t$-extended) Boltzmann measure, then
we average over the quenched distributions. Let us consider only
one party for simplicity: during the first operation we do not
take the whole party size but only a subsystem, say $k$ spins
(whose distribution is symmetric with respect to $0$ for both the
parties, $-1,+1$ for the dichotomic, Gaussians for the continuous
one). Then, in the second average, for any $k$ from $1$ to the
volume of the party, we consider all the possible links among
these $k$ nodes in this subgraph. As the links connecting the
nodes are always constant (i.e. equal to one due to the
Hubbard-Stratonovich transformation $(4.4)$) in the intensity, the
resulting associated energies are, in distribution and in the
thermodynamic limit, all equivalent: We are introducing a family
of microcanonical observables which sum up to a canonical one, in
some sense close to the decomposition introduced in
\cite{barraguerra}.

\subsection{The sum rule}

Let us now move on and consider the following  source $S$ of the
fluctuations of the order parameters, where
$\bar{M}_{l_b},\bar{N}_{l_c}$ stand for the replica symmetric
values\footnote{strictly speaking there are no replicas here but
configurations over different graphs. However the expression
RS-approximation, meaning that we assume the probability
distribution of the order parameters delta-like over their average
(denoted with a bar) is a sort of self-averaging and is an hinge
in disordered statistical mechanics such that we allow ourselves
to retain the same expression with a little abuse of language.} of
the previously introduced order parameters:
\begin{eqnarray}
S &=&
\left(\frac{1+a}{2}\right)\sqrt{\alpha\beta}\sum_{l_b}^{V-1}\sum_{l_c}^{L-1}P(l_b)P(l_c)\Big(
(M_{l_b}-\bar{M}_{l_b})(N_{l_c}-\bar{N}_{l_c})\Big)  \\
&=& \left(\frac{1+a}{2} \right)\sqrt{\alpha\beta}
 \langle \Big( M - \bar{M} \Big)\Big( N - \bar{N} \Big)
\rangle.
\end{eqnarray}
We see that with the choice of the parameters
$b_{l_c}=\sqrt{\alpha \beta}\bar{N}_{l_c}$ and
$c_{l_b}=\sqrt{\beta/\alpha}\bar{M}_{l_b}$, we can write the
$t$-streaming as
$$
\dot{A}=S - \frac{1+a}{2} \sqrt{\alpha \beta}\sum_{l_b}^{V-1}
\sum_{l_c}^{L-1} P(l_b)  P(l_c)\bar{M}_{l_b}\bar{N}_{l_c}.
$$
The replica symmetric solution (which is claimed to be the correct
expression in diluted ferromagnets) is simply achieved by setting
$S=0$ and forgetting it from future calculations.
\newline
We must now evaluate $A(0)$. This term is given by two separate
contributions, each for each party. Namely we have
\begin{eqnarray}\nonumber
A(0)&=& \frac1V \mathbb{E}\log\sum_{\sigma}e^{\sum_{l_c=1}^L
b_{l_c}\sum_i^V \eta_i \sigma_i} + \frac1V \mathbb{E} \log
\int_{-\infty}^{+\infty} \prod_{\mu=0}^L
d\mu(z_{\mu})e^{\sum_{l_b=1}^V c_{l_b} \sum_{\mu}^L
\chi_{\mu}z_{\mu}}
\\ \nonumber &=& \log 2 + \left(\frac{1+a}{2} \right) \sum_{l_b=0}^{V-1}P(l_b)
\sum_{l_c=0}^{L-1} P(l_c) \log\cosh\Big( \sqrt{\alpha\beta}
\bar{N}_{l_c}  \Big) + \left(\frac{1+a}{2} \right)^2 \frac{\beta}{2}
\sum_{l_b}^{V-1} P(l_b) \bar{M}_{l_b}^2.
\end{eqnarray}
Summing $A(0)$ plus the integral of $\partial_t[A(S=0)]$ we
finally get
\begin{eqnarray} A(t=1) &=& \log 2 + \left(\frac{1+a}{2} \right)
\sum_{l_c}^{L-1} \sum_{l_b}^{V-1} P(l_c) P(l_b)
\log\cosh\Big(\sqrt{\alpha \beta} \bar{N}_{l_c} \Big) \\ &+&
\frac{\beta}{2} \left(\frac{1+a}{2} \right)^2\sum_{l_b}^{L-1}P(l_b)\bar{M}_{l_b}^2
- \frac{1+a}{2}\sqrt{\alpha
\beta}\sum_{l_b}^{V-1}\sum_{l_c}^{L-1}P(l_b)P(l_c)\bar{M}_{l_b}\bar{N}_{l_c}.
\end{eqnarray}
It is possible to show that (as each bipartite ferromagnetic model
\cite{BGG}\cite{bg2}) the free energy obeys a min-max principle by
which, extremizing the free energy with respect to the order
parameters we can express $\langle \bar{N} \rangle$ trough
$\langle \bar{M} \rangle$: The trial replica symmetric solution,
expressed trough $\langle \bar{M}\rangle, \langle \bar{N}\rangle$
is (at fixed $\langle \bar{N} \rangle$) convex in $\langle \bar{M}
\rangle$. This defines uniquely
 a value $\langle \bar{M}(\bar{N})\rangle$ where we get the max. Further,
 $\langle \bar{M}(\bar{N})\rangle$ is increasing and convex in $\langle \bar{N}^2 \rangle$ such that
  the following extremization is a well defined procedure.
\begin{eqnarray}\label{self1}
\sum_k P(k) \frac{\partial A}{\partial \bar{M}_{k}} &=& 0 \to
\sum_{l_c}P(l_c)\bar{N}_{l_c} =  \sum_k P(k) \frac{1+a}{2} \sqrt{\frac{\beta}{\alpha}} \bar{M}_k, \\
\label{self2} \sum_k P(k) \frac{\partial A}{\partial
\bar{N}_{l_k}} &=& 0 \to \sum_{l_b}P(l_b)\bar{M}_{l_b}= \sum_k
P(k) \tanh\Big( \sqrt{\alpha \beta}\bar{N}_k\Big).
\end{eqnarray}
Due to the mean field nature of the model, as we can express
$\bar{N}_k$ trough the average of the $\bar{M}_k$, we can write
the free energy of our network trough the series of
$\bar{M}_{l_b}$ alone [as expected as we started by eq.
(\ref{modello})]
\begin{eqnarray}\label{soloM}
A(\beta,a) &=& \log 2 + (\frac{1+a}{2}) \log\cosh\Big(
\tanh^{-1}[\sum_{l_b}P(l_b)\bar{M}_{l_b}] \Big) \\ &+&
\frac{\beta}{2}(\frac{1+a}{2})^2\sum_{l_b}P(l_b)\bar{M}^2_{l_b} -
(\frac{1+a}{2})
\sum_{l_b}P(l_b)\bar{M}_{l_b}\tanh^{-1}\Big(\sum_{l_b'}P(l_{b'})\bar{M}_{l_b'}\Big).
\nonumber
\end{eqnarray}
As anticipated there is no true dependence by $\alpha$. Note that
without normalizing the scalar product among the bit strings we
should rescale $\beta$ accordingly with $\alpha$, as in the
$L\to\infty$ limit we would get an infinite coupling (which is
physically meaningless).
\newline
Before exploring further properties of these networks, we should
recover the well known limit of Curie-Weiss $a = +1$ and isolate
spin system $a = -1$.
\newline
Let us work out for the sake of clearness the self-consistency in
a purely Curie-Weiss style by extremizing, with respect to $\langle
\bar{M}, \rangle$ eq. (\ref{soloM}): \begin{eqnarray} \nonumber &&
\partial_{\langle \bar{M} \rangle}A(\beta) = (\frac{1+a}{2})\Big( \frac{\langle
\bar{M} \rangle}{1-\langle \bar{M} \rangle^2}  + \beta
(\frac{1+a}{2}) \langle \bar{M} \rangle \Big) -
(\frac{1+a}{2})\Big(\frac{\langle \bar{M} \rangle}{1-\langle
\bar{M} \rangle^2} + \tanh^{-1}\langle \bar{M}\rangle \Big) =0 \\
&& \Rightarrow \langle \bar{M} \rangle = \tanh(\beta
(\frac{1+a}{2}) \langle \bar{M} \rangle) \Rightarrow
\tanh^{-1}\langle \bar{M}\rangle=\beta (\frac{1+a}{2})\langle
\bar{M} \rangle,
\end{eqnarray} such that the to get the classical magnetization in
our model we have to sum overall the contributing graphs, namely
$\langle M_{CW} \rangle = \langle \bar{M} \rangle =
\sum_{l_b}P(l_b) \bar{M}_{l_b}$, and we immediately recover
\begin{eqnarray}\label{para}
a \to -1 &\Rightarrow& A(\beta, a=-1) = \log 2, \\ \label{ferro}
a=+1 &\Rightarrow& A(\beta, a =+1) = \log 2 + \log\cosh(\beta
\langle \bar{M} \rangle)-\frac{\beta}{2}\langle M^2 \rangle,
\end{eqnarray}
which are the correct limits (note that in eq. (\ref{para}) $J=0$,
while in eq. (\ref{ferro}) $J=1$).
\newline
Furthermore, we stress that in our ``topological microcanonical"
decomposition of our order parameters, when summing over all the
possible subgraphs to obtain the CW magnetization, these are all
null apart the only surviving of the fully connected network, so
the distribution of the order parameters becomes trivially
$\propto \delta(\bar{M}-M_{CW})$, namely, only one order parameter
survives, the classical Curie-Weiss magnetization.

\subsection{Critical line trough fluctuation theory}\label{fluttua}

Developing a fluctuation theory of the order parameters allows to
determine where critical behavior arises and, ultimately, the
existence of a phase transition\footnote{Strictly speaking this
 approach holds only for second order phase transition, which indeed is the one
expected in imitative models, even in presence of dilution
\cite{ABC}.}.
\newline
To this task we have at first to work out the general streaming
equation with respect to the $t$-flux. Given a generic observable
$\mathcal{O}$ defined on the space of the $\sigma,z$ variables, it
is immediate to check that the following relation holds (we set
$\alpha=1$ for the sake of simplicity as it never appears in the
calculations (as can be easily checked by substituting $\langle N
\rangle$ with $\langle M \rangle$ trough eq.~($4.23$) which
changes the prefactor from $(\frac{1+a}{2})\sqrt{\alpha \beta} \to
(\frac{1+a}{2})^2\beta$ and express the fluctuations only via the
real variables $\sigma$ \footnote{Another simple argument to
understand the useless of $\alpha$ is a comparison among neural
networks: in that context, $\alpha$ rules -in the thermodynamic
limit- the velocity by which we add stored memories into the
network with respect to the velocity by which we add neurons. If
the former are faster than a critical value, by a TLC argument
they sum up to a Gaussian before the infinite volume limit has
been achieved and the Hopfield model turns into an SK\cite{BGG}.
Here there is no danger in this as we have only positive
-normalized- interactions.}): \be \frac{\partial \langle
\mathcal{O}\rangle}{\partial t} =
\frac{1+a}{2}\sqrt{\beta}\Big[(\langle \mathcal{O}\mathcal{M}
\mathcal{N} \rangle - \langle \mathcal{O} \rangle \langle
\mathcal{M} \mathcal{N} \rangle ) - \bar{N}(\langle \mathcal{O}
\mathcal{M} \rangle - \langle \mathcal{O} \rangle \langle
\mathcal{M} \rangle) - \bar{M} (\langle \mathcal{O} \mathcal{N}
\rangle - \langle \mathcal{O} \rangle \langle \mathcal{N} \rangle
) \Big], \ee where we defined the centered and rescaled order
parameters:
\begin{eqnarray} \langle \mathcal{M} \rangle &=& \sqrt{V}\sum_{l_b}P(l_b)( M_{l_b} -
\bar{M}_{l_b}) =\sqrt{V}\langle M- \bar{M}\rangle, \\
\langle \mathcal{N} \rangle &=& \sqrt{L}\sum_{l_c}P(l_c)(N -
\bar{N}) = \sqrt{L}\langle N - \bar{N}\rangle. \end{eqnarray} Now
we focus on their squares: We want to obtain the behavior of
$\langle \mathcal{M}^2 \rangle_{t=1}$, $\langle \mathcal{M}
\mathcal{N} \rangle_{t=1}$, $\langle \mathcal{N}^2 \rangle_{t=1}$,
so to see where their divergencies (onsetting the phase
transition) are located.
\newline
By defining the dot operator as \be\label{dot} \langle
\dot{\mathcal{O}} \rangle = (\frac{1+a}{2})\sqrt{\beta}\partial_t
\langle \mathcal{O} \rangle \ee we can write
\begin{eqnarray}\nonumber
\dot{\langle \mathcal{M}^2 \rangle} &=& \Big[ \langle
\mathcal{M}^3 \mathcal{N} \rangle - \langle \mathcal{M}^2 \rangle
\langle \mathcal{M} \mathcal{N} \rangle - \bar{N}\langle
\mathcal{M}^3 \rangle + \bar{N} \langle \mathcal{M}^2 \rangle
\langle \mathcal{M} \rangle  - \bar{M}
\langle \mathcal{M}^2 \mathcal{N} \rangle + \bar{M} \langle \mathcal{M}^2 \mathcal{N} \rangle \Big], \\
\nonumber \dot{\langle \mathcal{M}\mathcal{N} \rangle} &=& \Big[
\langle \mathcal{M}^2 \mathcal{N}^2 \rangle - \langle \mathcal{M}
\mathcal{N} \rangle \langle \mathcal{M} \mathcal{N} \rangle -
\bar{N}(\langle \mathcal{M}^2 \mathcal{N} \rangle -  \langle
\mathcal{M} \mathcal{N} \rangle \langle \mathcal{M} \rangle)  -
\bar{M} (\langle \mathcal{M} \mathcal{N}^2 \rangle -  \langle
\mathcal{M} \mathcal{N} \rangle \langle \mathcal{N} \rangle)
\Big],
\\ \nonumber \dot{\langle \mathcal{N}^2 \rangle} &=& \Big[ \langle \mathcal{N}^3 \mathcal{M} \rangle -
\langle \mathcal{N}^2 \rangle \langle \mathcal{M} \mathcal{N}
\rangle - \bar{N}\langle \mathcal{N}^2 \mathcal{M} \rangle +
\bar{N} \langle \mathcal{N}^2 \rangle \langle \mathcal{M} \rangle
- \bar{M} \langle \mathcal{N}^3 \rangle + \bar{M} \langle
\mathcal{N}^2 \rangle \langle \mathcal{N} \rangle \Big].
\end{eqnarray}
Now, for the sake of simplicity, let us introduce alternative labels for the fundamental
observables. We define $A(t)=\langle
\mathcal{M}^2 \rangle_t$, $D(t)=\langle \mathcal{M} \mathcal{N}
\rangle_t$ and $G(t)= \langle \mathcal{N}^2 \rangle_t$ and let us
work out their $t=0$ value, which is straightforward as at $t=0$
everything is factorized (alternatively these can be seen as high
noise expectations):
$$
A(t=0) = 1, \ D(t=0) = 0, \
G(t=0)=\Big(1+(\frac{1+a}{2})^2\frac{\beta}{\alpha}\langle
\bar{M}^2 \rangle \Big)- \langle \bar{N}^2 \rangle = 1
$$
where we used the self-consistence relation ($4.23$) and assumed
that at least where everything is completely factorized the
replica solution is the true solution\footnote{Any debate
concerning RSB on diluted ferromagnets is however ruled out here
as we are approaching the critical line from above.}. Following
the technique introduced in \cite{sumrule}, starting from the high
temperature and, under the Gaussian ansatz for critical
fluctuations,  we want to take into account correlations among the
order parameters. Within this approach, using Wick theorem to
split the four observable averages in series of couples, the
 (formal) dynamical system reduces to
\begin{eqnarray}
\dot{A}(t) &=& 2 A(t) D(t), \\
\dot{D}(t) &=& A(t) G(t) + D^2(t), \\
\dot{G}(t) &=& 2 G(t) D(t).
\end{eqnarray}
We must now solve for $A(t),D(t),G(t)$ and evaluate these
expression at $t=\frac{1+a}{2}\sqrt{\beta}$ accordingly to the
definition of the dot operator in eq.(\ref{dot}). Notice at first
that
$$
\partial_t \log A = \frac{\dot{A}}{A}= 2D = \frac{\dot{G}}{G} =
\partial_t \log G.
$$
This means $\partial_t(A/G)=0$ and as $A(0)/G(0)=1$ we already
know that $A(t)=G(t)$: the fluctuations of the two order
parameters behave in the same way, not surprisingly, as already
pointed out their mutual interdependence several times.
\newline
We are left with
\begin{eqnarray}
\dot{D}(t) &=& G^2(t) + D^2(t), \\
\dot{G}(t) &=& 2G(t)D(t).
\end{eqnarray}
By defining $Y = D + G$ we immediately get, summing the two
equations above: $\dot{Y} = Y^2$ by which we get $Y(t)=
Y(0)/(1-tY(0))$. As $Y(0)=1$ we obtain that
$$
D(t=(\frac{1+a}{2})\sqrt{\beta}) +
G(t=(\frac{1+a}{2})\sqrt{\beta}) =
\frac{1}{1-(\frac{1+a}{2})\sqrt{ \beta}},
$$
so there is a regular behavior up to $\beta_c=
1/(\frac{1+a}{2})^2$.
\newline
We must now solve separately for $D$ and $G$: this is
straightforward by introducing the function $Z=G^{-1}$ and
checking that $Z$ obeys
$$
-\dot{Z} - 2YZ + 2=0,
$$
which, once solved with standard techniques (as $Y$ is known)
gives $G(t)=[2(1-t)]^{-1}$ and ultimately, simply by noticing the
divergencies of
$A(t=(1+a)\sqrt{\beta}/2),D(t=(1+a)\sqrt{\beta}/2),G(t=(1+a)\sqrt{\beta}/2)$,
we get the critical line for both the squared order parameters and
their relative correlation: All these functions do diverge on the
line \be \beta_c = \frac{1}{(\frac{1+a}{2})^2} = \frac{1}{ \bar{J}
}, \ee  defining a phase transition according with intuition.

\subsection{$L\to\infty$ scaling in the thermodynamic limit}

As we understood in Section $(3.3)$, in the $V\to\infty$ and
$L\to\infty$ limits we need to tune the limit of $a\to-1$
carefully to recover the various interesting topologies and to
avoid the trivial limits of fully connected/disconnected graph.

In particular, $a$ must approach $-1$ as $a=-1+\gamma/V^{\theta}$.
To tackle this scaling it is convenient to use directly
$\gamma,\theta$ as tunable parameter and rewrite the Hamiltonian
in the following forms\footnote{As we are going to see soon it is
not possible to normalize the Hamiltonian -both the coupling
strength and the volume extensiveness- for all the possible graphs
in only one expression. We choose to normalize so to tackle
immediately the better known limits, however apparent divergencies
in the couplings develop and can be standardly avoided by properly
rescaling the temperature corresponding to the amount of nearest
neighbors, as in more classical approaches.}
\begin{equation}
H_{V}(\sigma; \xi) = \frac{1}{2 \alpha
V^{2(1-\theta)}}\sum_{ij}^V\sum_{\mu}^L\xi_i^{\mu}\xi_j^{\mu}\sigma_i\sigma_j
\Rightarrow \tilde H_{V,L} (\sigma,z;\xi) =
\frac{\sqrt{\beta/\alpha}}{V^{1-\theta}}\sum_{i,\mu}\xi_{i,\mu}\sigma_i
z_{\mu},
\end{equation}
where the difference among the two expression
$H,\tilde H$ is due to the Hubbard-Stratonovick transformation
applied to the coupled partition functions, as performed early
trough eq.$(4.4)$.
\newline
Our free energy reads off now as \be
A(\alpha,\beta,\gamma,\theta)=\lim_{V \to \infty}\frac1V
\mathbb{E}\log\sum_{\sigma}\int\prod_{\mu}^L
d\mu(z_{\mu})\exp(\frac{\sqrt{\beta/\alpha}}{V^{(1-\theta)}})\sum_{i,\mu}\xi_{i
\mu}\sigma_i z_{\mu}, \ee where $\alpha$ accounts for the
different ratio among the two parties, $\beta$ the noise into the
network, $\theta$ selects the graph (see Sec.$(3.3)$) and $\gamma$
is the fine tuning inside the chosen topology.
\newline
The interpolating scheme remains the same: we introduce the right
amount of random fields and use $t\in(0,1)$ to define \be A(t) =
\frac1N \mathbb{E}\log\sum_{\sigma}\int d\mu(z_{\mu})\exp[t\tilde
H_{V,L}(\sigma,z;\xi) + (1-t)( \sum_i \tilde{a} \eta_i \sigma_i +
\sum_{\mu} \tilde{b} \eta_{\mu} z_{\mu})]. \ee By performing the
$t$-streaming easily we get \be
\partial_t A(t) = \frac{\sqrt{\beta} \gamma}{2}\langle (M - \bar
M)(N - \bar N) \rangle +
\frac{\sqrt{\beta}\gamma}{2}\bar{M}\bar{N}, \ee such that the
replica symmetric sum rule gets \be A(1) = A(0) -
\frac{\sqrt{\beta} \gamma}{2}\bar M \bar N, \ee and the replica
symmetric free energy reads off as \be A(\beta,\gamma,\theta) =
\log 2 + \frac{\gamma}{2V^{\theta}}\log\cosh(\sqrt{\beta}\bar N
V^{\theta}) + \frac{\beta \gamma^2}{8} \bar{M}^2 -
\frac{\sqrt{\beta} \gamma}{2}\bar M \bar N. \ee

Let us now investigate some limits of this expression and its
self-consistency. Note that by extremizing with respect to the
order parameters we can skip $\bar N$ trough $\bar M$ as \be \langle \bar
N \rangle = \frac{\sqrt{\beta}\gamma}{2}\langle \bar{M} \rangle. \ee

\subsubsection{$\theta=0$ case: Fully connected, weighted and Curie-Weiss scenario}

The case $\theta=0$ reduces to a fully connected graph, and in
particular in the upper bound for $\gamma$ (i.e. $\gamma=2$) its
topology recovers the unweighed CW model (see sec. $3.3$). We
should recover here even the CW thermodynamics.
\be A(\beta,\gamma,\theta=0)= \log 2 + \frac{\gamma}{2}\log\cosh(
\beta \frac{\gamma}{2} \langle \bar M \rangle) - \frac{ \beta
\gamma^2}{8} \langle \bar{M} \rangle^2, \ee
and its self-consistency relation reduces to
$$ \langle \bar M  \rangle = \tanh(\beta \frac{\gamma}{2} \langle  \bar{M}) \rangle.$$
This holds generally for the weighted  graph; in particular when
$\gamma = 2 \Rightarrow J = 1$ and the graph gets un-weighted
(still fully connected), we get the standard Curie-Weiss limit
once more:
\begin{eqnarray}
A(\beta,\gamma=2,\theta=0) &=& \log 2 + \log \cosh (\beta  \langle \bar M \rangle)
- {\beta}{2} \langle  \bar{M} \rangle^2, \\  \langle \bar{M}  \rangle &=& \tanh( \beta \langle  \bar{M} \rangle).
\end{eqnarray}

\subsubsection{$\theta = 1/2$: Standard dilution and Erdos-Renyi scenario}

With a scheme perfectly coherent with the previous one we can
write down free energy and its coupled self-consistency as
\begin{eqnarray}
A(\beta,\gamma,\theta=1/2) &=& \lim_{V \to \infty} \Big (\log 2 +
\frac{\gamma}{2\sqrt{V}}\log
\cosh (\frac{\beta \gamma}{2}\sqrt{V} \langle  \bar M \rangle) - \frac{\beta\gamma^2}{8} \langle  \bar{M} \rangle^2 \Big), \\
\langle \bar{M} \rangle &=& \lim_{V \to \infty} \tanh(\frac{\beta \gamma}{2}
\sqrt{V} \langle \bar{M} \rangle).
\end{eqnarray}
Let us stress that, as $\sqrt{J} = \frac{\gamma}{2\sqrt{V}}$, the argument of the logarithm of the hyperbolic cosine scales as $\sqrt{J}V\langle \bar M\rangle$:
This is coherent with the lack of a proper normalization into the Hamiltonian ($4.39$) because
for $\theta=1/2$ the latter is still divided by $V$ which should not appear.
To avoid the lack of a universal normalization, we need to renormalize the local average coupling by a factor $V$ so that we
get the correct behavior, namely we write explicitly the free energy putting in evidence that $p \sim 1 - \exp(-\alpha\gamma^2/4)$:
\be
A(\tilde\beta,\gamma,\theta=1/2) = \log 2 + \sqrt{J} \log\cosh( \frac{\tilde{\beta}\sqrt{J}}{p}\langle \bar M \rangle) - \frac{\tilde{\beta}J}{2p}\langle \bar M \rangle^2 , \ee
such that, being $\tilde \beta =  \beta p V $ \cite{ABC}, we can easily recover the trivial limits of the CW case when $p\to 1$ (and choerently $J\to1$, $\tilde\beta \to \beta$) and of the fully disconnected network $p\to 0 \Rightarrow A(\tilde\beta,\gamma\to 0, \theta=1/2) = \log 2$ as $p$ is superlinear in $\gamma$.

\subsubsection{$\theta = 1$: Extreme diluted regime}

With a scheme perfectly coherent with the previous one we can
write down free energy and its coupled self-consistency as
\begin{eqnarray}
A(\beta,\gamma,\theta=1) &=& \lim_{V \to \infty} \Big (\log 2 +
\frac{\gamma}{2V}\log
\cosh (\frac{\beta \gamma}{2}V  \langle \bar M \rangle)  - \frac{\beta\gamma^2}{8}\langle  \bar{M} \rangle^2 \Big), \\
\langle \bar{M} \rangle &=& \lim_{V\to \infty}\tanh(\frac{\beta \gamma}{2} V
\langle \bar{M} \rangle).
\end{eqnarray}
Of course here, with respect to the previous case, we get even stronger divergencies. Now we need to renormalize the local average coupling by a factor $V^2$.

\subsection{Numerics: Probability distribution}

As the critical line is obtained, in the fluctuation theory,
through the Gaussian ansatz, we double check our finding via
numerical simulations.

First of all, we notice that since the interaction matrix $J_{ij}$ is symmetric
$(J_{ij}=J_{ji})$, detailed balance holds and it is well known
\cite{ton1,capitolo} how to introduce a Markov process for
the dynamical evolution ruled by Hamiltonian (\ref{modello}) and
obtain the transition rates for stationarity: Montecarlo sampling
is then meaningful for equilibrium investigation.

The order parameter distribution function has been proved to be a
powerful tool for studying the critical line in different kinds of
systems; in particular, for magnetic systems the order parameter
can be chosen as the magnetization per spin which, in finite-size
systems, is a fluctuating quantity characterized by a probability
distribution $P(m)$ \cite{binder}. In Ising-like models undergoing
a second-order phase transition it is known that at temperatures
lower than the critical temperature $\beta^{-1}_c$, the
distribution $P(m)$ has a double peak, centered at the spontaneous
magnetization $+m$ and $-m$. At temperatures greater than
$\beta^{-1}_c$, $P(m)$ has a single peak at zero magnetization,
and exactly at $\beta^{-1}_c$ a double peak shape is observed.

In Fig.~\ref{fig:P_m} we plotted numerical  data for the
probability distribution, obtained by means of Monte Carlo
simulations, where $P(m)$ corresponds to the fraction of the total
number of realizations in which the system magnetization is $m$.
In the main figure we show the distribution for a system with
$V=1000$ set at a temperature $\beta^{-1}=1.1 \beta^{-1}_c$, while
in the inset we compare system of different sizes set at a
temperature $\beta^{-1} = 2 \beta^{-1}_c$. Notice that, for such
small temperatures, as the size in increased the distribution is
more and more peaked, while the probability to have zero
magnetization is vanishing; this corroborates the
replica-symmetric ansatz.


\begin{figure}[tb]
\includegraphics[height=80mm]{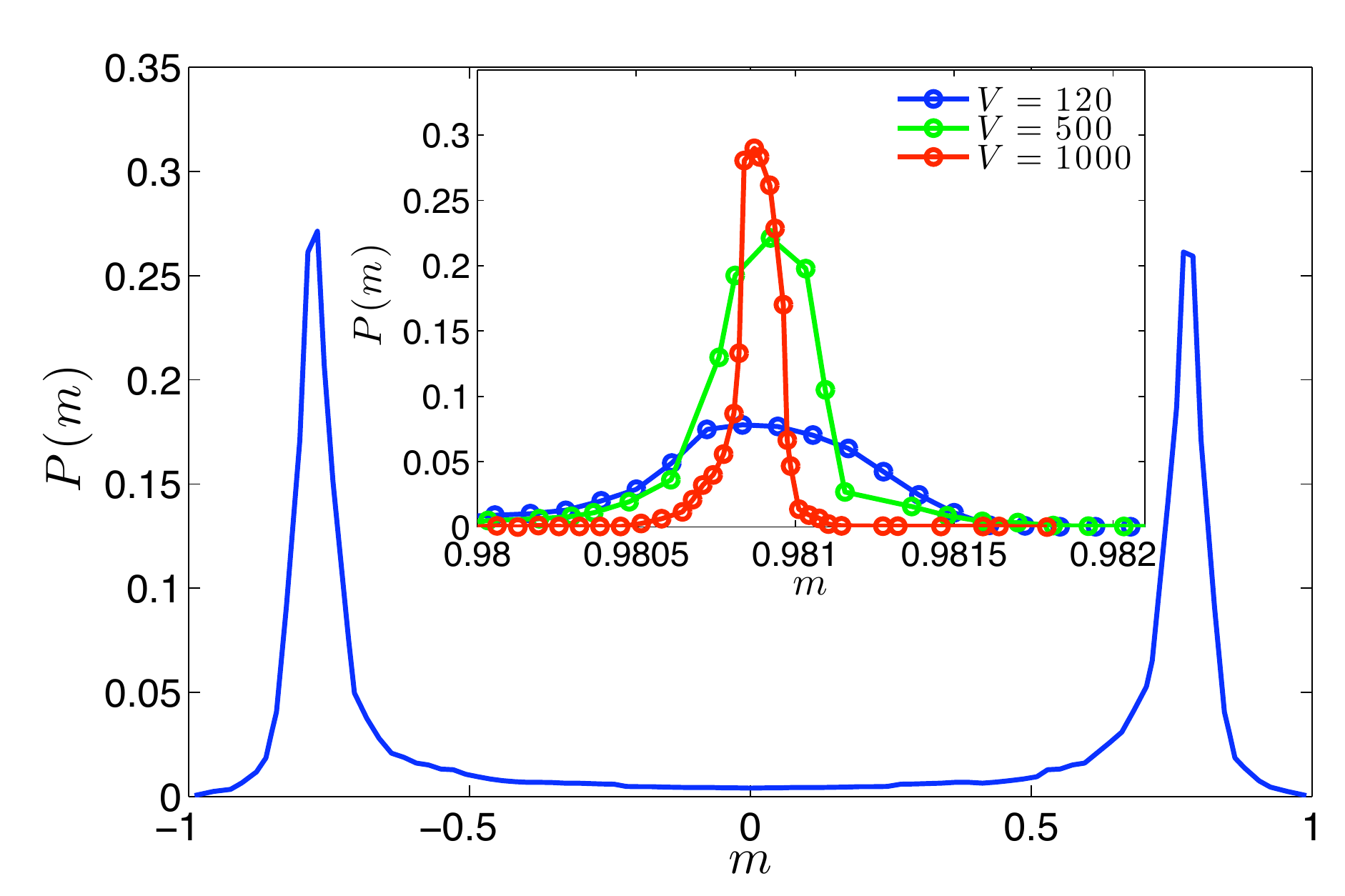}
\caption{\label{fig:P_m} Probability distribution for the order parameter $P(m)$. Main figure: system of size $N=1000$ at a temperature $T=0.5 T_c$; Inset: comparison between systems of size $N=120$, $N=500$ and $N=1000$, as shown by the legend, set at a temperature $T=0.9 T_c$.}
\end{figure}

\section{Conclusion}

In this paper we pioneered an alternative way  for obtaining
complex topologies. Interestingly from our approach small world
features are emergent properties and no longer imposed a-priori,
furthermore the core-theory descents from a simple shift $-1 \to
0$ in the definition domain of the patterns of an Hopfield model
and is able to recover all the best known complex topologies.
\newline
From a graph theory perspective we introduced a model which, given
a set of $V$ nodes, each corresponding to a set of $L$ attributes
encoded by a binary string $\xi$, defines an interaction coupling
$J_{ij} = (\xi_i \cdot \xi_j)$  for any couple of nodes $(i,j)$.
The resulting system can be envisaged by means of a weighted graph
displaying non trivial correlations among links. In particular,
when attributes are extracted according to a discrete uniform
distribution, i.e. $P(\xi_i^{\mu})=(1+a)/2$ for any $i \in [1, V]$
and $\mu \in [1,L]$, being $a$ a tunable parameter, we get that
when $a$ is sufficiently small the resulting network exhibits a
small-world nature, namely a large clustering coefficient; As $a$
is spanned, the network behaves as an isolate spin system, an
extreme dilute network, a linearly diverging connectivity network
a weighted fully connected and an un-weighted fully connected
network, respectively. Moreover, nodes are topologically
distinguishable according to the concentration $\rho$ of non-null
entries present in their corresponding binary strings:
interestingly, if the scaling among $L$ and $V$ is sub-linear
(i.e. $P \propto \ln V$ or even slower as in low storage networks \cite{agliba,baglia})
the degree distribution turns out to be multi-modal, each mode
pertaining to a different value of $\rho$. Instead, whenever the
scaling is (at least) linear -$L \propto V$-, the distribution
gets mono-modal. At least numerically, at finite $V,L$, when
looking at the distribution of weights, one finds that weak-ties
work as bridge, in full agreement with Granovetter theory: indeed,
one can detect small highly-connected clusters or "communities"
made up of nodes with similar attributes and links connecting
different communities are found to correspond a small coupling.
\newline
Then, as diluted models are of primary interest in disordered
statistical mechanics, by assuming self-averaging of the order
parameters, we solved the thermodynamic of the model: this
required a new technique (a generalization to infinitely random
fields of the double stochastic stability) which is of complete
generality as well and paves another way for approaching dilution
in complex systems.
\newline
Furthermore, within this framework, replicas are not necessary and
instead of averaging over these copies of the system (and dealing
with the corresponding overlap) we can obtain observables as
magnetization averages over local subgraphs, implicitly accounting
for a replica symmetric behavior (which is indeed assumed trough
the study).
\newline
An interesting finding, on which both the investigations converge
(graph theory and statistical mechanics), is a peculiar non-mean
field effect in the overall fields felt by the spins: from eq.
$(3.17)$ we see that the field insisting on a spin scales as
$\sqrt{J}$ while the averaged field on the network scales as $J$
-see eq. $(3.20)$- (which is the canonical mean field
expectation). Furthermore, looking at eq. $(4.25)$ we see that in
the hyperbolic tangent encoding the response of the spin to the
fields, the contribution of the other spins is not weighted by $J$
but by $\sqrt{J}$. As in the thermodynamics the coupling strength has been normalized,  $J<1 \to \sqrt{J}>J$: in complex
thermodynamics there is a super-linearity among the interactions:
despite this does not affect the critical behavior which is a
global feature of the system and consequently is found to scale
with $J$ (see eq. $(4.37)$), this may substantially change all the
other speculations based on intuition.
\newline
Of course, in the Curie-Weiss limit this effect disappear as global
and local environments do coincide (i.e. $J=1$).
\newline
It is worth stressing that (microscopic) correlation among
bit-strings is directly related to macroscopic behavior (e.g.
critical line), providing a new intriguing mechanics to study the
former via investigations on the latter (e.g. in social networks,
gene regulatory networks, or immune networks).
\newline
Next step in the research now should be double directed: from one
side, a clear statistical mechanics of scale free networks may
stem from our approach. From the other side, applications of this
theory to real systems (first at all a clear investigation on
dynamical retrieval properties), both in biology and in sociology,
should be a primary challenge as well.

\section*{Acknowledgments}

Francesco Guerra as usual is acknowledged for priceless scientific
and human interchange.
\newline
This work is supported by the FIRB grant: $RBFR08EKEV$
\newline
AB is grateful even to the Smart-Life Grant for partial support.
\newline
INFN and GNFM are acknowledged too for their partial support.

\end{document}